\documentclass[12pt]{article}
\usepackage{amsmath}
\usepackage{amssymb}

\newcommand{\bxx}{\mathbf{X}}
\newcommand{\bx}{\mathbf{x}}
\newcommand{\bpp}{\mathbf{P}}
\newcommand{\bp}{\mathbf{p}}
\newcommand{\tbeta}{\tilde{\beta}}
\newcommand{\bdp}{\mathbf{\Delta p}}
\newcommand{\bdx}{\mathbf{\Delta x}}
\newcommand{\bdpp}{\mathbf{\Delta P}}
\newcommand{\bdxx}{\mathbf{\Delta X}}
\newcommand{\thh}{\tilde{H}}

\title{
Composite system in deformed space with minimal length}
\author{C.\ Quesne$^{1}$ and V.\ M.\ Tkachuk$^{2}$\\ 
{\small\sl $^1$ Physique Nucl\'eaire Th\'eorique et Physique Math\'ematique,  Universit\'e Libre de Bruxelles,} \\ 
{\small\sl Campus de la Plaine CP229, Boulevard~du Triomphe, B-1050 Brussels, Belgium} \\
{\small \sl cquesne@ulb.ac.be} \\
{\small \sl $^2$ Department of Theoretical Physics, Ivan Franko National University of Lviv,}\\ 
{\small \sl 12 Drahomanov St., Lviv, UA-79005, Ukraine} \\
{\small \sl tkachuk@ktf.franko.lviv.ua}}
\date{ }
\begin{document}
\baselineskip=22pt plus 1pt minus 1pt
%%%%%%%%%%%%%%%%%%%%%%%%%%%%%%%%%%%%%%%%%%%%%%%%%%%%%%%%%%
\maketitle

\begin{abstract}
{}For composite systems made of $N$ different particles living in a space characterized by the same deformed Heisenberg algebra, but with different deformation parameters, we define the total momentum and the center-of-mass position to first order in the deformation parameters. Such operators satisfy the deformed algebra with new effective deformation parameters. As a consequence, a two-particle system can be reduced to a one-particle problem for the internal motion. As an example, the correction to the hydrogen atom $n$S energy levels is re-evaluated. Comparison with high-precision experimental data leads to an upper bound of the minimal length for the electron equal to $3.3\times 10^{-18} {\rm m}$. The effective Hamiltonian describing the center-of-mass motion of a macroscopic body in an external potential is also found. For such a motion, the effective deformation parameter is substantially reduced due to a factor $1/N^2$. This explains the strangely small result previously obtained for the minimal length from a comparison with the observed precession of the perihelion of Mercury. From our study, an upper bound of the minimal length for quarks equal to $2.4\times 10^{-17}{\rm m}$ is deduced, which appears close to that obtained for electrons.  
\end{abstract}

\vspace{0.5cm}

\vspace{0.5cm}

\newpage
\section{INTRODUCTION}

One of the important predictions of investigations in string theory and quantum gravity is the existence of a fundamental or minimal length following from the generalized uncertainty principle (GUP)
\begin{equation}
  \Delta X \ge \frac{\hbar}{2} \left( \frac{1}{\Delta P} + \beta \Delta P\right).  \label{eq:GUP}
\end{equation}
The latter implies a minimal position uncertainty $\Delta X_{\rm min} = \hbar \sqrt{\beta}$, which has been suggested to be of the order of the Planck length $l_p=\sqrt{\hbar G/c^3}\simeq 1.6\times 10^{-35}\rm m$.\par
%
%----------------------------------------------------------------------------------------------------------------------
%
In the simple one-dimensional case, the GUP can be derived from the deformed Heisenberg algebra with a small quadratic correction in momentum \cite{kempf95, kempf96}
\begin{equation}
  [X, P] = {\rm i} \hbar (1 + \beta P^2).  \label{eq:kempf-alg}
\end{equation}
One of the possible representations  of this algebra is provided by
\begin{equation}
  P = \frac{1}{\sqrt{\beta}} \tan (\sqrt{\beta} p), \qquad X = x,  \label{eq:kempf-rep}
\end{equation}
where $x$ and $p$ are the conventional position and momentum operators. To first order in $\beta$, such a representation can be written as
\begin{equation}
  P = (1 + \tfrac{1}{3} \beta p^2) p, \qquad X = x.  \label{eq:kempf-rep-lin}
\end{equation}
Although $X$ may be realized by an ordinary coordinate as in (\ref{eq:kempf-rep}) and (\ref{eq:kempf-rep-lin}), we have $\Delta X \ge \hbar \sqrt{\beta}$ because the quantum states for which $\Delta X < \hbar \sqrt{\beta}$ are nonphysical. As proved in \cite{kempf95}, the mean value of the kinetic energy in such states is indeed divergent, which is a property independent of the representation used for the algebra.\par 
%
%-------------------------------------------------------------------------------------------------
%
In three dimensions, a generalization of the deformed algebra (\ref{eq:kempf-alg}) reads \cite{kempf95, kempf97}
\begin{align}
  [X^{\mu}, P^{\nu}] &= {\rm i} \hbar [\delta^{\mu,\nu} (1 + \beta P^2) + \beta' P^{\mu} P^{\nu}], \qquad
        [P^{\mu}, P^{\nu}] = 0,  \label{eq:def-alg1}\\
  [X^{\mu}, X^{\nu}] &= {\rm i} \hbar \frac{(2\beta - \beta') + (2\beta + \beta') \beta P^2}{1 + \beta P^2}
        (P^{\mu} X^{\nu} - P^{\nu} X^{\mu}),  \label{eq:def-alg2}
\end{align}
where $\mu$, $\nu=1$, 2, 3 (or $x$, $y$, $z$), and $\beta$, $\beta'$ are two deformation parameters which are assumed positive ($\beta$, $\beta' \ge 0$). The minimal length now becomes $\Delta X_{\rm min} = \hbar \sqrt{\beta + \beta'}$.\par 
%
%-------------------------------------------------------------------------------------------------------------------
% 
In this respect, it is worth pointing out that a long time ago a deformed algebra leading to a quantized spacetime with a natural unit of length was already introduced by Snyder in a relativistic context \cite{snyder}\par
%
%-----------------------------------------------------------------------------------------------------------
%
It follows from (\ref{eq:def-alg2}) that in general the configuration space becomes nonuniform, namely it is not covariant under translation of spatial coordinates anymore. In the present paper, we consider the special case $\beta' = 2 \beta$, wherein the position operators commute to first order in $\beta$ and the configuration space is uniform again. In such a linear approximation, the deformed Heisenberg algebra reads \cite{brau}
\begin{equation}
  [X^{\mu}, P^{\nu}] = {\rm i} \hbar [\delta^{\mu,\nu} (1 + \beta P^2) + 2\beta P^{\mu} P^{\nu}], \qquad
  [P^{\mu}, P^{\nu}] = [X^{\mu}, X^{\nu}] = 0.  \label{eq:def-alg-lin}
\end{equation}
As it can be easily checked, the Jacobi identity, which was satisfied by the algebra (\ref{eq:def-alg1}), (\ref{eq:def-alg2}), is still valid for (\ref{eq:def-alg-lin}) to first order in $\beta$. Furthermore, the minimal length, which is a consequence of the commutation relations between position and momentum operators, but not of those between different position operators, is still nonvanishing (for another recent example, see \cite{ali}). It can be found by considering, for instance, the commutator of $X^1$ with $P^1$, which from (\ref{eq:def-alg-lin}) is given by
\begin{equation}
  [X^1, P^1] = {\rm i} \hbar \{1 + 3 \beta (P^1)^2 + \beta [(P^2)^2 + (P^3)^2]\}.
\end{equation}
The Heisenberg uncertainty relation now reads
\begin{equation}
\begin{split}
  \Delta X^1 \Delta P^1 & \ge \frac{\hbar}{2} \{1 + 3 \beta \langle (P^1)^2 \rangle  + \beta 
       [\langle (P^2)^2 \rangle + \langle (P^3)^2 \rangle]\} \\
  & \ge \frac{\hbar}{2} [1 + 3 \beta \langle (P^1)^2 \rangle]  \\
  & \ge \frac{\hbar}{2} [1 + 3 \beta (\Delta P^1)^2],  \label{eq:Heisenberg}
\end{split}
\end{equation}
so that 
\begin{equation}
  \Delta X^1 \ge \frac{\hbar}{2} \left( \frac{1}{\Delta P^1} + 3 \beta \Delta P^1\right),
\end{equation}
and similarly for the other components. Comparing with (\ref{eq:GUP}) directly leads to the minimal length $\Delta X_{\rm min} = \hbar \sqrt{3 \beta}$, in agreement with the general result for the algebra (\ref{eq:def-alg1}), (\ref{eq:def-alg2}).\par 
%
%-------------------------------------------------------------------------------------------------------------------
% 
The properties of an algebra being independent of the representation considered, it is convenient to use the simplest one, which for (\ref{eq:def-alg-lin}) reads
\begin{equation}
  P^{\mu} = (1 + \beta p^2) p^{\mu}, \qquad X^{\mu} = x^{\mu},  \label{eq:rep}
\end{equation}
in terms of the conventional momentum and position operators $p^{\mu}$, $x^{\mu}$, satisfying the (nondeformed) canonical commutation relations
\begin{equation}
  [x^{\mu}, p^{\mu}] = {\rm i} \hbar \delta^{\mu,\nu}, \qquad [p^{\mu}, p^{\nu}] = [x^{\mu}, x^{\nu}] = 0.
\end{equation}
In the same linear approximation, the inverse transformation reads
\begin{equation}
  p^{\mu} = (1 - \beta P^2) P^{\mu}, \qquad x^{\mu} = X^{\mu}.
\end{equation}
It is worth stressing that Eq.\ (\ref{eq:rep}) is but a generalization to three dimensions of the one-dimensional relation (\ref{eq:kempf-rep-lin}).\par
%
%-----------------------------------------------------------------------------------------------------------
%
The observation that the GUP can be obtained from a deformed algebra has opened the possibility of studying the influence of the minimal length on physical properties of several single-particle systems on the quantum level as well as on the classical one. In the classical limit, the commutator of quantum mechanical operators is replaced by the Poisson bracket of the corresponding classical variables.\par
%
%-------------------------------------------------------------------------------------------------------------------
%
Deformed commutation relations bring new difficulties in quantum mechanics as well as in classical one. As a consequence, only a few problems are known for which the energy spectra have been found exactly. These are the one-dimensional harmonic oscillator with minimal uncertainty in position \cite{kempf95} and also with minimal uncertainty in both position and momentum \cite{cq03, cq04}, the $D$-dimensional isotropic harmonic oscillator \cite{chang, dadic}, the three-dimensional Dirac oscillator \cite{cq05}, and the one-dimensional Coulomb problem \cite{fityo}. Note that a $D$-dimensional isotropic harmonic oscillator was studied for the first time in \cite{kempf97} using perturbation theory over the deformation parameters. A Lorentz-covariant deformed algebra with minimal length and its application to the $(1+1)$-dimensional Dirac oscillator were considered in Ref.~\cite{cq06}. The three-dimensional Coulomb problem with a deformed Heisenberg algebra was studied in a perturbation theory framework \cite{brau, benczik05, stetsko06a, stetsko06b, stetsko08}. Classical mechanics with deformed Poisson brackets was studied in Refs.~\cite{benczik02, frydryszak, silagadze}. Recently, the influence of a minimal length on the Lamb shift, the Landau levels, and the tunneling current in a scanning tunneling microscope was examined \cite{das}. In Ref.~\cite{vakili}, the effects of noncommutativity and of the existence of a minimal length on the phase space of a cosmological model were investigated. The authors of Ref.~\cite{battisti} analysed several physical consequences following from the noncommutative Snyder spacetime geometry.\par
%
%------------------------------------------------------------------------------------------------------------------
% 
All of these problems correspond to one-particle systems in a deformed space with minimal length. As it has been observed elsewhere \cite{kowalski04, kowalski05}, a satisfactory treatment of many-particle systems in such a context has not been achieved so far. In particular, the way of defining the total momentum and the center of mass has remained unclear. The purpose of this paper is to fill in this gap in the nonrelativistic case. We actually plan to study composite systems made of $N$ different particles living in a deformed space charaterized by the same deformed Heisenberg algebra, but with different deformation parameters, and to compare some predictions coming from our refined model with those previously made using cruder approximations.\par
%
%----------------------------------------------------------------------------------------
% 
In Sec.~II, we review the two-body problem in a deformed space. We then study in Sec.~III the influence of the minimal length on the hydrogen atom considered as a two-particle system. A generalization to the $N$-body problem is considered in Sec.~IV and is applied in Sec.~V to a macroscopic body, corresponding to the $N \to \infty$ limit, with an estimation of the minimal length from the observation of planetory motion. Finally, Sec.~VI contains the conclusion.\par
%
%================================================================
%
\section{TWO-BODY PROBLEM IN DEFORMED SPACE}

It is well known that in the case of two interacting particles in ordinary space, one can introduce external degrees of freedom, the total momentum and the center-of-mass position, and internal ones, the relative momentum and  position. After separation of the external and internal degrees of freedom, the two-body problem can be reduced to a one-body problem. In this Section, we plan to show how to carry out a similar separation in the case of a deformed space.\par
%
%-------------------------------------------------------------------------------------------------------
%
Let us assume that in deformed space the Hamiltonian has a similar form as in nondeformed one, which means that in the absence of external potential it can be written as
\begin{equation}
  H_2 = \frac{P_1^2}{2m_1} + \frac{P_2^2}{2m_2} + V(|\bxx_1 - \bxx_2|), \label{eq:H-2}
\end{equation}
where $V(|\bxx_1 - \bxx_2|)$ is the interaction potential energy of the two particles. In the general case, the different particles of masses $m_i$ may feel different (effective) deformations so that the operators $X_1^{\mu}$ and $P_1^{\mu}$ satisfy the deformed algebra (\ref{eq:def-alg-lin}) with some deformation parameter $\beta_1$, while the operators $X_2^{\mu}$ and $P_2^{\mu}$ fulfil the same algebra but with a different deformation parameter $\beta_2$. It is also natural to suppose that the operators corresponding to different particles commute with one another.\par
%
%-------------------------------------------------------------------------------------------------------------------
%
The main problem to solve is how to define the total momentum and the center-of-mass position of the two-particle system. The central idea of this paper is to introduce the total momentum in deformed space as an integral of motion, then to define the center-of-mass position as its conjugate variable. For such a purpose, let us rewrite the Hamiltonian (\ref{eq:H-2}) in nondeformed space by using the representation (\ref{eq:rep}). The result reads
\begin{equation}
  H_2 = \frac{p_1^2}{2m_1} (1 + 2 \beta_1 p_1^2) + \frac{p_2^2}{2m_2} (1 + 2 \beta_2 p_2^2) 
  + V(|\bx_1 - \bx_2|), \label{eq:H-2bis}
\end{equation}
where we note that only the kinetic energy operator gets deformed. It is easy to find that $\bp_1 + \bp_2$ commutes with $H_2$ and is therefore an integral of motion, which we can associate with the total momentum $\bp_0 = \bp_1 + \bp_2$ in nondeformed space. In such a space, its conjugate operator is the center-of-mass position $\bx_0$ and we can also introduce the operators $\bx$, $\bp$, describing the relative motion, in the traditional way,
\begin{align}
  &\bx_0 = \mu_1 \bx_1 + \mu_2 \bx_2, & &\bp_0 = \bp_1 + \bp_2, \label{eq:x_0}  \\
  &\bx = \bx_1 - \bx_2, & &\bp = \mu_2 \bp_1 - \mu_1 \bp_2,  \label{eq:x}
\end{align}  
the pairs $(\bx_0, \bp_0)$ and $(\bx, \bp)$ satisfying the canonical Heisenberg algebra. In (\ref{eq:x_0}) and (\ref{eq:x}), we have defined $\mu_1 = m_1/(m_1 + m_2)$ and $\mu_2 = m_2/(m_1 + m_2)$. The inverse transformation reads
\begin{align}
  \bx_1 &= \bx_0 + \mu_2 \bx, & \bp_1 &= \mu_1 \bp_0 + \bp,   \label{eq:x-1} \\
  \bx_2 &= \bx_0 - \mu_1 \bx, & \bp_2 &= \mu_2 \bp_0 - \bp.  \label{eq:x-2}
\end{align}  
\par
%
%------------------------------------------------------------------------------------------------------------
%
Substituting Eqs.~(\ref{eq:x-1}) and (\ref{eq:x-2}) into the two-particle kinetic energy operator (the first two terms of (\ref{eq:H-2bis})), we obtain
\begin{equation}
  T_2 = T_0 + T + \Delta T  \label{eq:T-2}
\end{equation}
with
\begin{gather}
  T_0 = \frac{p_0^2}{2m} [1 + 2 (\beta_1 \mu_1^3 + \beta_2 \mu_2^3) p_0^2],  \label{eq:T-0}  \\
  T = \frac{p^2}{2\mu} [1 + 2 (\beta_1 \mu_2 + \beta_2 \mu_1) p^2],  \\
  \begin{split}
    \Delta T &= \frac{1}{m} \{(\beta_1 \mu_1 + \beta_2 \mu_2) [4 (\bp_0 \cdot \bp)^2 + 2 p_0^2 p^2]
           + 4 (\beta_1 \mu_1^2 - \beta_2 \mu_2^2) p_0^2 (\bp_0 \cdot \bp) \\
    &\quad + 4 (\beta_1 - \beta_2) p^2 (\bp_0 \cdot \bp)\},   
  \end{split}  \label{eq:Delta-T}
\end{gather}
where we have introduced  the total mass $m = m_1 + m_2$ and the reduced mass $\mu = m_1 m_2/(m_1 + m_2) = m_1 \mu_2 = m_2 \mu_1$. The first two contributions $T_0$ and $T$ to $T_2$, depending only on the (nondeformed) total momentum and relative momentum, respectively, may be considered as the kinetic energy of the center of mass and that of the relative motion. In contrast to what happens in nondeformed space, $T_2$ also contains an additional term $\Delta T$, proportional to the deformation parameters and describing the influence of the relative motion on the center-of-mass one or vice versa.\par
%
%-----------------------------------------------------------------------------------------------------------
%
On comparing (\ref{eq:T-0}) with either of the first two terms on the right-hand side of (\ref{eq:H-2bis}), we see that $T_0$ has the form of a kinetic energy operator in deformed space with a new deformation parameter
\begin{equation}
  \tbeta_0 = \beta_1 \mu_1^3 + \beta_2 \mu_2^3.  \label{eq:cm-beta}
\end{equation}
Introducing then the total momentum in deformed space with this deformed parameter $\tbeta_0$,
\begin{equation}
  P_0^{\mu} = (1 + \tbeta_0 p_0^2) p_0^{\mu},
\end{equation}
as in (\ref{eq:rep}), we find
\begin{equation}
  T_0 = \frac{P_0^2}{2m}.
\end{equation}
\par
%
%----------------------------------------------------------------------------------------------------------------
%
Similarly, the second term of (\ref{eq:T-2}) can be rewritten in the form 
\begin{equation}
  T = \frac{P^2}{2\mu}, \qquad P^{\mu} = (1 + \tbeta p^2) p^{\mu},
\end{equation}
where the deformation parameter for the relative motion is
\begin{equation}
  \tbeta = \beta_1 \mu_2 + \beta_2 \mu_1  \label{eq:rm-beta}
\end{equation}
and differs from that for the center-of-mass motion given in (\ref{eq:cm-beta}).\par
%
%----------------------------------------------------------------------------------------------------------------
%
Since the third term $\Delta T$ of (\ref{eq:T-2}) is proportional to the deformation parameters, to lowest order in $\beta_1$, $\beta_2$, we may replace the lower-case $\bp_0$, $\bp$ by the capital ones $\bpp_0$, $\bpp$ on the right-hand side of (\ref{eq:Delta-T}). As a result, the total kinetic energy operator $T_2$ has been expressed in terms of the deformed total and relative momenta, $\bpp_0$, $\bpp$.\par
%
%---------------------------------------------------------------------------------------------------------------
% 
To $\bpp_0$ and $\bpp$, we can now associate the corresponding position operators
\begin{equation}
  \bxx_0 = \bx_0, \qquad \bxx = \bx,
\end{equation}
respectively. The operators for the center-of-mass motion (resp.\ the relative motion), $X_0^{\mu}$ and $P_0^{\mu}$ (resp. $X^{\mu}$ and $P^{\mu}$), satisfy Eq.~(\ref{eq:def-alg-lin}) with $\beta$ replaced by $\tbeta_0$ (resp.\ $\tbeta$), while $X_0^{\mu}$ and $P_0^{\mu}$ commute with $X^{\mu}$ and $P^{\mu}$.\par
%
%-----------------------------------------------------------------------------------------------------------
% 
The Hamiltonian (\ref{eq:H-2bis}) now becomes
\begin{equation}
  H_2 = \frac{P_0^2}{2m} + \frac{P^2}{2\mu} + \Delta T + V(|\bxx|).
\end{equation}
Since the total momentum $\bpp_0$ commutes with it, each of its components is an integral of motion, which may be replaced by its eigenvalue, so that we are left with an eigenvalue problem for the relative motion. The two-body problem has therefore been reduced to a one-body problem as in the conventional space, but it is important to stress that the center-of-mass motion now influences the relative one through the presence of $\bpp_0$ in $\Delta T$.\par
%
%----------------------------------------------------------------------------------------------------------------------
%
It is worth noting that, in general, the Hamiltonian may contain an extra contribution coming from some external field. In the case where the external potential changes very slowly on distances of the order of the system size, we may assume that the external potential only depends on $\bxx_0$, so that the total Hamiltonian reads
\begin{equation}
  H_2 = \frac{P_0^2}{2m} + \frac{P^2}{2\mu} + \Delta T + V(|\bxx|) + V_{\rm ext}(\bxx_0).
\end{equation}
\par
%
%======================================================================
%
\section{EFFECT OF THE MINIMAL LENGTH ON THE HYDROGEN ATOM ENERGY SPECTRUM}

The hydrogen atom may be considered as a two-particle system, for which particle 1 is the proton and particle 2 the electron. Hence in this case, $m_1 = m_{\rm p}$, $\beta_1 = \beta_{\rm p}$, $m_2 = m_{\rm e}$, and $\beta_2 = \beta_{\rm e}$. In the absence of external field, the hydrogen atom Hamiltonian reads
\begin{equation}
  H_2 = \frac{P_0^2}{2m} + \frac{P^2}{2\mu} + \Delta T - \frac{e^2}{|\bxx|}
\end{equation}
and the components of the total momentum are integrals of motion.\par
%
%----------------------------------------------------------------------------------------------------------
% 
Let us replace the latter by their eigenvalues, which we denote by the same symbols. The eigenvalue problem for $H_2$ is then equivalent to 
\begin{equation}
  H_{\rm a} \psi = E' \psi, \qquad E' = E - \frac{P_0^2}{2m},
\end{equation}
where the hydrogen atom Hamiltonian in deformed space reads
\begin{equation}
  H_{\rm a} = \frac{P^2}{2\mu} + \Delta T - \frac{e^2}{|\bxx|}
\end{equation}
and contains a term $\Delta T$ depending on the value of the total momentum. To first order in the deformation parameters, $H_{\rm a}$ can be written in canonical (nondeformed) space as
\begin{equation}
  H_{\rm a} = \frac{p^2}{2\mu} - \frac{e^2}{|\bx|} + \tbeta \frac{p^4}{\mu} + \Delta T, 
\end{equation}
where $\tbeta (p^4/\mu) + \Delta T$ is a small correction taking into account the deformation of the commutation relations.\par
%
%-------------------------------------------------------------------------------------------------------------------
%
The effect of such a correction on the $nS$ energy level can be easily calculated in the first order of perturbation theory and leads to the additional contribution
\begin{equation}
  \Delta E_n = \Delta E_{n,1} + \Delta E_{n,2},  \label{eq:Delta-E}
\end{equation}
where
\begin{gather}
  \Delta E_{n,1} = \left\langle n, 0, 0 \left| \tbeta \frac{p^4}{\mu} \right| n, 0, 0\right\rangle = 2 \tbeta
       \frac{8n-3}{n^2} \frac{\hbar^2}{a^2} |E_n^0| = 2 \tbeta (\mu c)^2 \alpha^2 \frac{8n-3}{n^2} 
       |E_n^0|, \label{eq:Delta-E-1}  \\
  \begin{split}  
  \Delta E_{n,2} &= \langle n, 0, 0 | \Delta T | n, 0, 0\rangle = \frac{20}{3} (\beta_1 \mu_1 + \beta_2 
       \mu_2) \frac{\mu}{m} p_0^2 |E_n^0| \\
  &= \frac{20}{3} (\beta_1 \mu_1 + \beta_2 \mu_2) \frac{m}{\mu} (\mu c)^2 \left(\frac{v}{c}\right)^2 
       |E_n^0|.
  \end{split}  \label{eq:Delta-E-2}     
\end{gather}
Here $a = \hbar^2/(\mu c^2)$ is the Bohr radius, $\alpha = e^2/(\hbar c) \simeq 1/137$ is the fine-structure constant, $p_0 = mv$, where $v$ is the hydrogen atom velocity, and $E_n^0 = - e^2/(2an^2)$ is the unperturbed energy. The first term on the right-hand side of (\ref{eq:Delta-E}) reproduces the result obtained for the first time \cite{brau} with the simplifying assumption $\tbeta = \beta_1 \mu_2 + \beta_2 \mu_1 \to \beta$. The second term is new and describes the influence of the center-of-mass motion on the relative one in deformed space.\par
%
%-----------------------------------------------------------------------------------------------------------------------------
%
On taking into account that $m_{\rm e}/m_{\rm p} \simeq 1/1840$, we find
\begin{equation}
  \mu_1 \simeq 1, \qquad \mu_2 \simeq 1/1840, \qquad \mu \simeq m_{\rm e}.
\end{equation}
On the other hand, it is natural to suppose that the deformation parameters for elementary particles, such as the electron and the quarks, are approximately the same, {\sl i.e.}, $\beta_{\rm e} \simeq \beta_{\rm q}$. Since the proton is made of three quarks, it follows from Eq.~(\ref{eq:cm-beta-ter}), to be proved in Sect.~IV, that the deformation parameter for the proton is
\begin{equation}
  \beta_{\rm p} = \frac{\beta_{\rm q}}{3^2} \simeq \frac{\beta_{\rm e}}{3^2}, 
\end{equation}
provided we assume that the effective (constituent) mass of the quarks in the proton is the same. Then, for the hydrogen atom, the effective deformation parameters for the center-of-mass and the relative motions, given in Eqs.~(\ref{eq:cm-beta}) and (\ref{eq:rm-beta}), respectively, become
\begin{equation}
  \tbeta_0 \simeq \beta_{\rm p} + \beta_{\rm e} \left(\frac{1}{1840}\right)^3 \simeq \beta_{\rm p},
\end{equation}
and 
\begin{equation}
  \tbeta \simeq \beta_{\rm p} \frac{1}{1840} + \beta_{\rm e} \simeq \beta_{\rm e}.
\end{equation}
Similarly, the factor containing deformation parameters in (\ref{eq:Delta-E-2}) reduces to
\begin{equation}
  \beta_1 \mu_1 + \beta_2 \mu_2 \simeq \beta_{\rm p} + \beta_{\rm e} \frac{1}{1840} \simeq \beta_{\rm p}.
\end{equation}
\par
%
%-------------------------------------------------------------------------------------------------------------------------
%
We conclude that for the hydrogen atom, the deformation parameter for the center-of-mass motion is determined by the proton and that for the relative motion by the electron. Furthermore, the two corrections $\Delta E_{n,1}$ and $\Delta E_{n,2}$ to the energy spectrum have a different origin, the former being proportional to $\beta_{\rm e}$ and caused by the electron motion and the latter being proportional to $\beta_{\rm p}$ and brought about by the proton motion.\par
%
%--------------------------------------------------------------------------------------------------------------
%
Let us now compare the contributions of these two terms to the energy spectrum. For this purpose, let us consider their ratio
\begin{equation}
  \frac{\Delta E_{n,2}}{\Delta E_{n,1}} = \frac{10}{3} \frac{n^2}{8n-3} \frac{\beta_{\rm p}}{\beta_{\rm e}}
  \frac{m_{\rm p}}{m_{\rm e}} \frac{1}{\alpha^2} \left(\frac{v}{c}\right)^2 \simeq 1.3 \times 10^7
  \frac{n^2}{8n-3} \left(\frac{v}{c}\right)^2.  
\end{equation}
For the ground level, we find equal contributions, {\sl i.e.}, $\Delta E_{1,2} / \Delta E_{1,1} = 1$, when the hydrogen atom velocity is $2 \times 10^5 {\rm m/s}$, this estimate remaining of the same order for the first excited level. Such a velocity corresponds to a rather high temperature, namely $T = 1.6 \times 10^6 {\rm K}$.\par
%
%---------------------------------------------------------------------------------------------------------------------------
%
{}Following Ref.~\cite{brau}, to estimate an upper bound of the minimal length we use the experiment of high-precision spectrometry of the 1S--2S two-photon transition in atomic hydrogen. Recently, some significant progress has been achieved in the accuracy of the 1S--2S frequency measurement, which has now reached the level of $1.4 \times 10^{-14}$ \cite{kolachevsky}. In this experiment, the hydrogen was cooled to the temperature $T \simeq 5 {\rm K}$ and the velocity of the atomic beam used for the measurement was $v \simeq 120 {\rm m/s}$. In such a case, the second term $\Delta E_{n,2}$ may be neglected in comparison with the first one $\Delta E_{n,1}$.\par
%
%----------------------------------------------------------------------------------------------------------
%
Let us determine the correction $\Delta_{12} = \Delta E_2 - \Delta E_1 = \Delta E_{2,1} - \Delta E_{1,1}$ to the energy $E_{12}^0 = E_2^0 - E_1^0$ of the 1S--2S transition. From (\ref{eq:Delta-E-1}), it follows that
\begin{equation}
  \frac{\Delta_{12}}{E_{12}^0} = - \frac{67}{6} \frac{\hbar^2 \beta_{\rm e}}{a^2}.
\end{equation}
On assuming that the effect of the minimal length on the energy spectrum cannot be seen experimentally yet, we find
\begin{equation}
  \frac{|\Delta_{12}|}{E_{12}^0} < 1.4 \times 10^{-14}.
\end{equation}
This leads to the upper bound of the deformation parameter for the electron 
\begin{equation}
  \hbar \sqrt{\beta_{\rm e}} \le  1.9 \times 10^{-18} {\rm m},
\end{equation}
from which follows the upper bound
\begin{equation}
  \Delta X_{\rm min}^{\rm e} = \hbar \sqrt{3\beta_{\rm e}} \le  3.3 \times 10^{-18} {\rm m}  \label{eq:upper} 
\end{equation}
of the minimal length in the deformed space wherein the electron is living.\par
%
%------------------------------------------------------------------------------------------------------------------
%
Observe that we used some experimental data corresponding to more accurate measurements than those employed in Ref.~\cite{brau}. Our result for the upper bound of the minimal length is therefore one order less than the previous one. It is also worth pointing out that the authors of Refs.~\cite{benczik05, stetsko06a, stetsko06b} based their estimation of the minimal length on another argument related to the discrepancy between theoretical and experimental values of the Lamb shift for the hydrogen atom $n$S levels. On assuming that such a discrepancy can be entirely attributed to the minimal length correction to the energy spectrum, it was found that the upper bound of the minimal length ranges from $10^{-16} \rm m$ to $10^{-17} \rm m$ for different values of the deformation parameters. For the case $2\beta = \beta'$ considered here, it was shown in Ref.~\cite{stetsko06a} that the upper bound is $1.6 \times 10^{-16} \rm m$. So we may state that the result contained in (\ref{eq:upper}) for the upper bound of the minimal length is somewhat smaller than that found in previous papers.\par
%
%===============================================================
%  
\section{\boldmath $N$-BODY PROBLEM IN DEFORMED SPACE}

The purpose of this Section is to introduce internal and external degrees of freedom for $N$ particles in deformed space by generalizing what has been done for two particles in Sect.~II.\par
%
%-------------------------------------------------------------------------------------------------------------------
%
Let us assume again that the $N$-particle Hamiltonian in deformed space has a similar form as in nondeformed one. In the absence of external potential, it can therefore be written as
\begin{equation}
  H_N = \sum_i \frac{P_i^2}{2m_i} + \frac{1}{2} \sum_{\substack{i, j \\ i \ne j}} V(|\bxx_i - \bxx_j|),
\end{equation}
where $i$ and $j$ run over 1, 2, \ldots, $N$, the position and momentum operators $X_i^{\mu}$, $P_i^{\mu}$ satisfy the commutation relations (\ref{eq:def-alg-lin}) with some deformation parameter $\beta_i$, and the operators corresponding to different particles commute with one another. According to (\ref{eq:rep}), these operators can be represented by
\begin{equation}
  P_i^{\mu} = (1 + \beta_i p_i^2) p_i^{\mu}, \qquad X_i^{\mu} = x_i^{\mu},
\end{equation}
in terms of canonical position and momentum operators $x_i^{\mu}$, $p_i^{\mu}$.\par
%
%-----------------------------------------------------------------------------------------------------------
%
In nondeformed space, we can introduce external and internal degrees of freedom associated with
\begin{equation}
  \bp_0 = \sum_i \bp_i, \qquad \bx_0 = \sum_i \mu_i \bx_i,
\end{equation}
and
\begin{equation}
  \bdp_i = \bp_i - \mu_i \bp_0, \qquad \bdx_i = \bx_i - \bx_0,
\end{equation}
respectively. Here $\mu_i = m_i/\sum_j m_j$. It can be easily checked that the former operators commute with the latter and that the relative motion operators satisfy the conditions
\begin{equation}
  \sum_i \bdp_i = 0, \qquad \sum_i \mu_i \bdx_i = 0,
\end{equation}
showing that there are in fact $3(N-1)$ internal degrees of freedom, as it should be. Note that the total momentum in nondeformed space $\bp_0$ commutes with the Hamiltonian $H_N$, rewritten in terms of the canonical operators.\par
%
%-----------------------------------------------------------------------------------------------------
%
As for two particles, the kinetic energy operator  can be re-expressed as
\begin{equation}
  T_N = \frac{P_0^2}{2m} + \sum_i \frac{\Delta P_i^2}{2m_i} + \Delta T,  \label{eq:T-N}
\end{equation}
where
\begin{equation}
  \Delta T = \frac{1}{m} \sum_i \beta_i [2\mu_i P_0^2 \Delta P_i^2 + 4\mu_i (\bpp_0 \cdot \bdpp_i)^2
  + 4\mu_i^2 P_0^2 (\bpp_0 \cdot \bdpp_i) + 4 \Delta P_i^2 (\bpp_0 \cdot \bdpp_i)].  \label{eq:Delta-Tbis}
\end{equation}
In (\ref{eq:T-N}), the total and relative momentum operators in deformed space read
\begin{equation}
  P_0^{\mu} = (1 + \tbeta_0 p_0^2) p_0^{\mu}, \qquad \Delta P_i^{\mu} = (1 + \beta \Delta p_i^2)
  \Delta p_i^{\mu}.
\end{equation}
The former, together with the conjugate position operators $X_0^{\mu} = x_0^{\mu}$, satisfy a deformed algebra of type  (\ref{eq:def-alg-lin}) with an effective deformation parameter
\begin{equation}
  \tbeta_0 = \sum_i \beta_i \mu_i^3.  \label{eq:cm-beta-bis}
\end{equation}
The first term in (\ref{eq:T-N}) is the kinetic energy of the center-of-mass motion with total mass $m = \sum_i m_i$, the second term is that of the relative motion, while the third term is only present in deformed space and describes the influence of the relative motion on the center-of-mass one or vice versa. Note that for two particles, since $\bdp_1 = - \bdp_2 = \bp$, Eq.~(\ref{eq:T-N}) reproduces the result (\ref{eq:T-2}) obtained in Sec.~2. For the total Hamiltonian $H_N$, the components of $\bpp_0$ are integrals of motion.\par
%
%-------------------------------------------------------------------------------------------------------------------------
%
In the presence of an external field varying very slowly on distances of the order of the system size, $H_N$ reads
\begin{equation}
  H_N = H_0 + H + \Delta T,
\end{equation}
where
\begin{gather}
  H_0 = \frac{P_0^2}{2m} + V_{\rm ext}(\bxx_0), \\
  H = \sum_i \frac{\Delta P_i^2}{2m_i} + \frac{1}{2} \sum_{\substack{i, j \\ i \ne j}} V(|\bdxx_i - \bdxx_j|),
\end{gather}
and $\Delta T$ is given in (\ref{eq:Delta-Tbis}). Here $H_0$ and $H$ describe the external and internal motions, respectively, while $\Delta T$ leads to their entanglement due to the deformation.\par
%
%----------------------------------------------------------------------------------------------------------------
%
Before concluding this Section, let us mention two interesting properties of the effective deformation parameter $\tbeta_0$.\par
%
%----------------------------------------------------------------------------------------------------------
% 
{}First, let us divide  the composite system into two subsystems corresponding to $i=1$, 2, \ldots, $N_1$ and $i = N_1 + 1$, $N_1 + 2$, \ldots, $N$, respectively. If we calculate the effective deformation parameter for each subsystem separately, {\sl i.e.},
\begin{equation}
  \tbeta_1 = \sum_{i=1}^{N_1} \beta_i \left(\frac{m_i}{\sum_{j=1}^{N_1} m_j}\right)^3, \qquad
  \tbeta_2 = \sum_{i=N_1+1}^{N} \beta_i \left(\frac{m_i}{\sum_{j=N_1+1}^{N} m_j}\right)^3,
\end{equation}
then the effective deformation parameter of the whole system reads
\begin{equation}
  \tbeta_0 = \tbeta_1 \left(\frac{\sum_{i=1}^{N_1} m_i}{\sum_{j=1}^{N} m_j}\right)^3 + \tbeta_2 
  \left(\frac{\sum_{i=N_1+1}^{N} m_i}{\sum_{j=1}^{N} m_j}\right)^3,
\end{equation}
in agreement with Eq.~(\ref{eq:cm-beta-bis}).\par
%
%-----------------------------------------------------------------------------------------------------------------------
%
Second, if the system is made of $N$ particles with the same masses $m_1 = m_2 = \cdots = m_N = m$ and the same deformation parameters $\beta_1 = \beta_2 = \cdots = \beta_N = \beta$, then its effective deformation parameter is given by
\begin{equation}
  \tbeta_0 = \frac{\beta}{N^2}.  \label{eq:cm-beta-ter}
\end{equation}
In the case of a composite system consisting of a large number of particles, one therefore observes a substantial reduction of $\tbeta_0$ with respect to the deformation parameter $\beta$ for the individual particles. In other words, though the latter may have a certain fixed minimal uncertainty in position, their collective center of mass can be localized better and better the more particles the system has. This might be interpreted as a noise reduction due to an averaging effect.\par
%
%=============================================================
%
\section{MACROSCOPIC BODY IN DEFORMED SPACE AND MINIMAL LENGTH ESTIMATION FROM THE OBSERVATION OF PLANETARY MOTION}

In the present Section, we consider a macroscopic body in deformed space, corresponding to $N \to \infty$, and study the motion of its center of mass in an external potential.\par
%
%--------------------------------------------------------------------------------------------------------------
%
{}For a macroscopic system, the external degrees of freedom change in time more slowly than the internal ones. It is natural to assume that the system of particles is in its equilibrium state and that the internal degrees of freedom are distributed according to $\exp(-H/k_{\rm B}T)/Z$, where $Z$ is the partition function, $T$ the temperature, and $k_{\rm B}$ the Boltzmann's constant. Then averaging over the internal degrees of freedom, we obtain an effective Hamiltonian
\begin{equation}
  \langle H_N \rangle = H_0 + \langle \Delta T \rangle + \langle H \rangle = \thh_0 + \langle H \rangle,
\end{equation}
describing the center-of-mass motion. Here $\langle H \rangle$ is the internal energy of the system, which is a function of the temperature and does not depend on the external degrees of freedom. Hence it may be treated as a constant and $\thh_0 = H_0 + \langle \Delta T \rangle$ may be considered as the effective Hamiltonian describing the center-of-mass motion.\par
%
%-------------------------------------------------------------------------------------------------------------
% 
To calculate the average of $\Delta T$, defined in (\ref{eq:Delta-Tbis}), let us use the fact that the mean value of the product of an odd number of internal momenta vanishes, hence
\begin{equation}
  \langle \bpp_0 \cdot \bdpp_i \rangle = \langle \Delta P_i^2 (\bpp_0 \cdot \bdpp_i) \rangle = 0,
\end{equation}
and the property
\begin{equation}
  \langle \Delta P_i^{\mu} \Delta P_i^{\nu} \rangle = \delta^{\mu,\nu} \tfrac{1}{3} \langle \Delta P_i^2
  \rangle,
\end{equation}
together with the relation between the mean kinetic energy and the temperature,
\begin{equation}
  \langle \Delta P_i^2 \rangle = 2m_i \left\langle \frac{\Delta P_i^2}{2m_i} \right\rangle =
  3m_i k_{\rm B} T.
\end{equation}
Note that since $\Delta T$ is proportional to the deformation parameters, in all these relations, the deformed momenta may be replaced by nondeformed ones. The result for $\langle \Delta T \rangle$ reads
\begin{equation}
  \langle\Delta T \rangle = 10 k_{\rm B} T P_0^2 \sum_i \beta_i \mu_i^2. 
\end{equation}
\par
%
%-----------------------------------------------------------------------------------------------------------
%
{} Finally, the effective Hamiltonian describing the center-of-mass motion can be written in the form
\begin{equation}
  \thh_0 = \frac{P_0^2}{2m^*} + V_{\rm ext}(\bxx_0),
\end{equation}
where $X_0^{\mu}$, $P_0^{\mu}$ satisfy the deformed Heisenberg algebra (\ref{eq:def-alg-lin}) with the effective deformation parameter defined in (\ref{eq:cm-beta-bis}) and
\begin{equation}
  m^* = \frac{m}{1 + 20 k_{\rm B} T m \sum_i \beta_i \mu_i^2}
\end{equation}
is an effective mass.\par
%
%-------------------------------------------------------------------------------------------------------------------
%
To describe the motion of a macroscopic body, we can use the classical limit, wherein the commutators of quantum mechanical operators are replaced by the Poisson brackets of the corresponding classical variables via $[\ldots, \ldots]/({\rm i}\hbar) \Rightarrow \{\ldots, \ldots\}$. As a result, we get classical mechanics with deformed Poisson brackets
\begin{equation}
\begin{split}
  \{F, G\} &= \sum_{\mu,\nu} \left(\frac{\partial F}{\partial X^{\mu}} \frac{\partial G}{\partial P^{\nu}} -
        \frac{\partial F}{\partial P^{\mu}} \frac{\partial G}{\partial X^{\nu}}\right) \{X^{\mu}, P^{\nu}\} \\
  &\quad + \sum_{\mu,\nu} \left(\frac{\partial F}{\partial X^{\mu}} \frac{\partial G}{\partial X^{\nu}} 
        \{X^{\mu}, X^{\nu}\} + \frac{\partial F}{\partial P^{\mu}} \frac{\partial G}{\partial P^{\nu}} 
        \{P^{\mu}, P^{\nu}\}\right), \label{eq:def-Poisson}
\end{split}
\end{equation}
where the fundamental Poisson brackets are determined from the corresponding commutation relations. In our case, for algebra (\ref{eq:def-alg-lin}), we have
\begin{equation}
  \{X^{\mu}, P^{\nu}\} = \delta^{\mu,\nu} (1 + \beta P^2) + 2\beta P^{\mu} P^{\nu}, \qquad  
  \{X^{\mu}, X^{\nu}\} = \{P^{\mu}, P^{\nu}\} = 0.
\end{equation}
\par
%
%----------------------------------------------------------------------------------------------------------------
%
The time evolution of the mechanical system is governed by the Hamilton equations
\begin{equation}
  \dot{X}^{\mu} = \{X^{\mu}, H\}, \qquad \dot{P}^{\mu} = \{P^{\mu}, H\} 
\end{equation}
with the deformed Poisson brackets defined in (\ref{eq:def-Poisson}). For details on classical mechanics with deformed Poisson brackets see, {\sl e.g.}, Refs.~\cite{benczik02, frydryszak}.\par
%
%-------------------------------------------------------------------------------------------------------------------------------
%
In Ref.~\cite{benczik02}, the authors studied the effect of the deformation produced by the algebra (\ref{eq:def-alg1}), (\ref{eq:def-alg2}), which also includes (\ref{eq:def-alg-lin}) as a special case, on the classical orbits of particles in a central force potential. To lowest order in the deformation parameters they derived the correction to the precession angle of a planet caused by deformation. Comparing their result to the observed precession  of the perihelion of Mercury, they estimated an upper bound of the deformation parameter or of the minimal length,
\begin{equation}
  \hbar \sqrt{\beta} < 2.3 \times 10^{-68} {\rm m}.  \label{eq:benczik}
\end{equation}
\par
%
%-----------------------------------------------------------------------------------------------------------------------
%
This strangely small result, 33 orders of magnitude below the Planck length, was obtained because of the implicit assumption made by the authors that the deformation parameters for Mercury are the same as for elementary particles. As shown above, for a composite system made of elementary particles, such as Mercury, one should use, instead of the elementary particle deformation parameter $\beta$, the effective deformation parameter $\tbeta_0$, defined in Eq.~(\ref{eq:cm-beta-bis}), so that Eq.~(\ref{eq:benczik}) should be replaced by
\begin{equation}
  \hbar \sqrt{\tbeta_0} < 2.3 \times 10^{-68} {\rm m}.  \label{eq:new-benczik}
\end{equation}
\par
%
%-------------------------------------------------------------------------------------------------------------------
%
To determine the value of $\tbeta_0$ for Mercury, let us first estimate the number of elementary particles contained in the planet. Since the main contribution to its mass comes from the nucleons (neutrons and protons), from the mass of Mercury $M = 3.3 \times 10^{23} {\rm kg}$ and that of nucleons $m_{\rm nuc} = 1.67 \times 10^{-27} {\rm kg}$, we find for the number of nucleons in Mercury $N_{\rm nuc} = 2 \times 10^{50}$. Moreover, the number of electrons is equal to the number of protons, which is approximately half the number of nucleons, {\sl i.e.}, $N_{\rm e} = N_{\rm p} \simeq N_{\rm nuc}/2 \simeq 10^{50}$. On using (\ref{eq:cm-beta-bis}), we can now relate the effective deformation parameter for Mercury to the deformation parameters for nucleons $\beta_{\rm nuc}$ and electrons $\beta_{\rm e}$,
\begin{equation}
  \tbeta_0 = N_{\rm nuc} \beta_{\rm nuc} \left(\frac{m_{\rm nuc}}{M}\right)^3 + N_{\rm e} \beta_{\rm e}   
  \left(\frac{m_{\rm e}}{M}\right)^3, \label{eq:def-Mercury}
\end{equation}
where we assume that protons and neutrons have the same deformation parameters $\beta_{\rm p} = \beta_{\rm n} = \beta_{\rm nuc}$. Furthermore, it is easy to estimate that
\begin{equation}
  \frac{m_{\rm nuc}}{M} \simeq \frac{1}{N_{\rm nuc}}, \qquad \frac{m_{\rm e}}{M} \simeq \frac{m_{\rm e}}
  {N_{\rm nuc}m_{\rm nuc}} \simeq \frac{1}{1840 N_{\rm nuc}}.
\end{equation}
On the other hand, since the nucleons are made of three quarks, Eq.~(\ref{eq:cm-beta-ter}) leads to $\beta_{\rm nuc} = \beta_{\rm q}/3^2$. On assuming the same deformation parameters for elementary particles such as electrons and quarks, $\beta_{\rm e} = \beta_{\rm q}$, we therefore find that the second term in (\ref{eq:def-Mercury}) may be omitted and hence
\begin{equation}
  \tbeta_0 = \frac{\beta_{\rm nuc}}{N_{\rm nuc}^2}.
\end{equation}
\par
%
%-------------------------------------------------------------------------------------------------------------
%
Substituting this value of $\tbeta_0$ into (\ref{eq:new-benczik}), we obtain
\begin{equation}
  \hbar \sqrt{\beta_{\rm nuc}} < N_{\rm nuc} \times 2.3 \times 10^{-68} {\rm m} = 4.6 \times 10^{-18} 
  {\rm m}
\end{equation}
and for the minimal length for nucleons
\begin{equation}
  \Delta X^{\rm nuc}_{\rm min} = \hbar \sqrt{3\beta_{\rm nuc}} < 8.0 \times 10^{-18} {\rm m}.
\end{equation}
For quarks, the upper bounds of the deformation parameter and of the minimal length are given by
\begin{equation}
  \hbar \sqrt{\beta_{\rm q}} = 3 \hbar \sqrt{\beta_{\rm nuc}} < 1.4 \times 10^{-17} {\rm m}
\end{equation}
and
\begin{equation}
  \Delta X^{\rm q}_{\rm min} = 3 \Delta X^{\rm nuc}_{\rm min} < 2.4 \times 10^{-17} {\rm m},
\end{equation}
respectively. It is worth stressing that these upper bounds are substantially higher than those found in Ref.~\cite{benczik02} and are close to those obtained in Sec.~III for the electron although the experimental data used in both cases come for completely different types of measurements.\par
%
%====================================================================
%
\section{CONCLUSION}

In the present paper, we have proposed for the first time a consistent description of many-particle systems in a deformed space with minimal length. On assuming that the $N$ particles making up the composite system live in a space characterized by the same deformed Heisenberg algebra, but with different deformation parameters, we have been able to define, to first-order in the deformation parameters, the total momentum and the center-of-mass position, as well as the relative momenta and coordinates.\par
%
%-----------------------------------------------------------------------------------------------------------
%
In contrast to ordinary space, the external and internal operators are not separated in the Hamiltonian, hence the internal motion has an influence on the external one and vice versa. Since in the absence of external potential, however, the total momentum is conserved, its components may be replaced by their eigenvalues when considering the eigenvalue problem for $N$ particles in deformed space. By proceeding in this way, a two-particle problem, for instance, can still be reduced to a one-particle problem for the internal motion.\par
%
%-------------------------------------------------------------------------------------------------------------------
% 
Another important aspect of the definition of the external and internal degrees of freedom is that the corresponding operators satisfy the same deformed commutation relations as the individual particle operators, but with different effective deformation parameters. In particular, for a system of identical particles, the effective deformation parameter for the center-of-mass motion turns out to be given by $\tbeta_0 = \beta/N^2$, which means a substantial reduction for large $N$.\par
%
%-------------------------------------------------------------------------------------------------------------------------
% 
{}Furthermore, we have found the effective Hamiltonian describing the centre-of-mass motion of a macroscopic body (corresponding to the $N \to \infty$ limit) in an external potential. Such a motion can be described in the framework of classical mechanics provided standard Poisson brackets are replaced by deformed ones.\par
%
%-------------------------------------------------------------------------------------------------------------------
% 
As a first application of our new formalism, we have determined the first-order correction to the $n$S energy levels of the hydrogen atom, considered as a two-particle system. Such a correction contains a term due to the internal motion, which is proportional to the deformation parameter for the electron and reproduces previous theoretical estimates \cite{brau, benczik05, stetsko06a} in some limit. In the present case, however, there also appears a new contribution due to the influence of the center-of-mass motion on the relative one and determined by the deformation parameter for the proton, as well as by the atom squared velocity $v^2$. This second term becomes comparable to the first one for a rather large velocity or, in other words, for a rather high temperature $T$.\par
%
%-----------------------------------------------------------------------------------------------------------
%
We have then compared our results with data coming from a recent experiment of high-precision spectrometry of the 1S--2S two-photon transition in atomic hydrogen \cite{kolachevsky}. Since, in the latter, $v$ and $T$ are low, our second correction term to the energy levels is negligeable. The first one has provided us with an upper bound of the minimal length for the electron equal to $\Delta X_{\rm min}^{\rm e} = \hbar \sqrt{3\beta_{\rm e}} \le  3.3 \times 10^{-18} {\rm m}$. Such a bound is approximately one order less than that obtained in Ref.~\cite{brau}, due to the gain in experimental accuracy recently made.\par
%
%-----------------------------------------------------------------------------------------------------------------------
%
{}Finally, we have re-examined the estimation of the minimal length upper bound, which was made in Ref.~\cite{benczik02} by comparing the calculated correction to the precession of the perihelion of Mercury with the observed one and which had led to $\hbar \sqrt{\beta} < 2.3 \times 10^{-68} {\rm m}$. We have pointed out that this strangely small result (33 orders of magnitude below the Planck length) is entirely due to the use of the individual particle deformation parameter $\beta$ instead of the effective deformation parameter $\tbeta_0$ for Mercury motion and to the large number of elementary particles making up the planet. From the inequality $\hbar \sqrt{\tbeta_0} < 2.3 \times 10^{-68} {\rm m}$, we have got an upper bound of the minimal length for quarks given by $\Delta X^{\rm q}_{\rm min} < 2.4 \times 10^{-17} {\rm m}$.\par
%
%------------------------------------------------------------------------------------------------------------
%
In conclusion, we would like to stress that it makes sense comparing upper bounds for different elementary particles, but not for composite systems made of different numbers of elementary particles. It is remarkable that the upper bounds  found here for electrons and quarks are rather similar although they come from completely different measurements, related to the hydrogen atom spectrum and planetory motion, respectively. It should be noted, however, that such bounds are still rather far from the Planck scale.\par
%
%----------------------------------------------------------------------------------------------------------
%
As a last remark, we would like to mention another application of the framework developed in the present paper. As it was first noticed in \cite{kempf97}, deformed commutation relations not only allow one to study the consequences of a possible fundamental minimal length for elementary particles, but also provide an effective approach to the finiteness of composite particles. In this spirit, one might model the extendedness of the proton in the hydrogen atom (while possibly keeping the electron pointlike) and compare the predicted effects to those resulting from a description of the proton in terms of quarks and to experimental data. We think that this is an interesting open problem for future investigation. 
%
%==================================================================
%
\section*{ACKNOWLEDGMENTS}

V.M.T. thanks the National Fund for Scientific Research (FNRS), Belgium, for financial support during his stay
at Universit\'e Libre de Bruxelles.
\par
%
%=================================================================
%

\end{document}